\documentstyle[12pt,epsf]{article}
\newcommand{\thspace}{\kern.08333em}

\def \beq{\begin{equation}}
\def \eeq{\end{equation}}
\def \beqn{\begin{eqnarray}}
\def \eeqn{\end{eqnarray}}
\def \Bbar{\bar B}
\def \Abar{\bar A}
\def \Dbar{\bar{D}^0}
\def \Kbar{\bar{K}^0}
\def \bbar{\bar{b}}
\def \ubar{\bar {u}}
\def \cbar{\bar {c}}
\def \sbar{\bar{s}}
\def \s{\sqrt{2}}

\def \beq{\begin{equation}}
\def \eeq{\end{equation}}
\def \beqn{\begin{eqnarray}}
\def \eeqn{\end{eqnarray}}
\def \b{{\cal B}}
\def \Dbar{\bar{D}^0}
\def \Kbar{\bar{K}^0}
\def \Bbar{\bar B}
\def \Tbar{\bar T}
\def \Cbar{\bar C}
\def \abar{\bar{a}}
\def \bbar{\bar{b}}
\def \ubar{\bar {u}}
\def \cbar{\bar {c}}
\def \sbar{\bar{s}}
\def \s{\sqrt{2}}

\begin{document}
\rightline{FERMILAB-PUB-98/227-T}
\rightline{EFI-98-29}
\rightline{hep-ph/9807447}
\bigskip
\bigskip
\centerline{\bf FINAL STATE INTERACTION EFFECTS ON $\gamma$ FROM $B \to D 
K$}
\bigskip
\centerline{\it Michael Gronau\footnote{Permanent Address: Physics 
Department,
Technion -- Israel Institute of Technology, 32000 Haifa, Israel.}} 
\centerline{\it Fermi National Accelerator Laboratory}
\centerline{\it P. O. Box 500, Batavia, IL 60510}
%\midskip
\centerline {and}
\centerline{\it Jonathan L. Rosner}
\centerline{\it Enrico Fermi Institute and Department of Physics}
\centerline{\it University of Chicago, Chicago, IL 60637}
\vskip 1cm  

\centerline{\bf ABSTRACT}
\bigskip

\begin{quote}
The implications of a negligible annihilation contribution in $B \to D 
K$ decays are reanalyzed and are shown to lead to no new constraints on 
the weak phase
$\gamma$ from color-allowed $B^{\pm} \to D K^{\pm}$ decays. A test of  negligible annihilation is proposed in $B^+ \to D^+ K^0$ (or $B^+ \to D^+ K^{*0}$), and 
an application is presented in which $\gamma$ can be determined from 
these processes (or corresponding $B \to D K^*$ decays) supplemented 
with isospin-related neutral $B$ decays.

\end{quote}
\bigskip

\leftline{\qquad PACS codes:  12.15.Hh, 12.15.Ji, 13.25.Hw, 14.40.Nd}
\bigskip

Recently one of us proposed a method \cite{MG} to constrain the CKM weak 
phase $\gamma =$ $-{\rm Arg}(V^*_{ub}V_{ud}/V^*_{cb}V_{cd})$ from 
{\it color-allowed} $B^{\pm} \to D K^{\pm}$ decays \cite{previous}, when 
both 
flavor and CP-eigenstate neutral $D$ mesons are considered.
Decays with flavor states have already been observed by CLEO \cite{CLEO} 
with branching ratio $\b(B^{\pm} \to D K^{\pm})/\b(B^{\pm} \to D 
\pi^{\pm})  
= 0.055 \pm 0.014 \pm 0.005$, or $\b(B^{\pm} \to D K^{\pm}) \simeq 3 
\times 10^{-4}$. Our approach was
general and involved no dynamical assumptions about hadronic weak matrix
elements and about final state interactions. 
Subsequently Xing \cite{Xing} claimed that a certain improvement in this 
method may be achieved by making the dynamical assumption of a negligible annihilation contribution. 
While this assumption is reasonable, it may be spoiled by rescattering 
effects \cite{BGR, rescat} and would have to be tested experimentally. 
One of the purposes of the present letter is to suggest such a test. Our 
second purpose is to 
go over the arguments in \cite{Xing} and to point out a certain flaw in 
the treatment of final state interactions. By presenting a correct 
analysis we 
will show that, in fact, the assumption of a vanishing annihilation contribution
does not lead to any further constraint on $\gamma$ beyond the one 
obtained 
in \cite{MG}. Finally, we will present a scheme \cite{HoKo} which 
involves 
also {\it neutral} $B$ decays to $DK$ states through which $\gamma$ 
can be determined when neglecting annihilation. 
Since these decay modes involve $B^0 \to D K^0$, in which the neutral 
$B$ 
meson must be flavor-tagged, it would be advantageous to consider 
instead 
the corresponding self-tagged decays $B \to D K^*$. All the 
considerations applied below to $B \to D K$ apply also to $B \to D K^*$. 

For completeness, let us recapitulate the results of \cite{MG}. 
Writing
\beq
A(B^+ \to \Dbar K^+) = \abar e^{i\bar{\Delta}}~~,~~~~~
A(B^+ \to D^0 K^+) = a e^{i\Delta} e^{i\gamma}~~,
\eeq
and introducing the two CP-eigenstates, $D_{1,2}=(D^0 \pm \Dbar)/\s$,
one considers the two charge-averaged ratios of rates for these states
and for the flavor states
\beq\label{R_i}
R_i \equiv \frac{2[\Gamma(B^+ \to D_i K^+) + \Gamma(B^- \to D_i K^-)]}
{\Gamma(B^+ \to \Dbar K^+) + \Gamma(B^- \to D^0 K^-)}~~,~~~~~i = 1,2~~.
\eeq
One finds
\beq\label{R12}
R_{1,2} = 1 + r^2 \pm 2r\cos\delta\cos\gamma~~,
\eeq
where $r \equiv a/\abar, \delta \equiv \Delta-\bar{\Delta}$. 
This leads to two inequalities
\beq\label{LIMIT}
\sin^2\gamma \leq R_{1,2}~~,~~~~~i = 1,2~~,
\eeq
which could potentially imply new constraints on $\gamma$ in future 
experiments \cite{MG}. 

The two pseudo-asymmetries
\beq\label{A_i}
{\cal A}_i \equiv \frac{\Gamma(B^+ \to D_i K^+) - \Gamma(B^- \to D_i 
K^-)}
{\Gamma(B^+ \to \Dbar K^+) + \Gamma(B^- \to D^0 K^-)}~~,~~~~~i = 1,2~,
\eeq
are given by
\beq
{\cal A}_2 = -{\cal A}_1 = r \sin\delta \sin\gamma~~,
\eeq 
and together with the two ratios $R_i$ could, in principle, provide 
sufficient information to determine the three parameters $r, \delta$ 
and 
$\gamma$ (up to certain discrete ambiguities). However, since $r$ is 
suppressed by a smaller than one ratio of CKM factors and by a color 
factor, one expects $r \approx 0.1$, which would be too small to be 
measured from the tiny deviation of $(R_1  + R_2)/2$ from unity. 
Similarly, unless $\delta$ is very large, the asymmetries may be too 
small to permit nonzero measurements. 

While the above equations and constraints follow generally from the CKM structure of the weak charged currents, one may try to supplement these equations with assumptions about the dynamics of the above hadronic 
decays. 
One such common assumption \cite{SU3} is the neglect of annihilation 
diagrams. This assumption was made
in \cite{Xing}, where it was claimed to reduce the number of independent parameters by essentially relating $r$ and $\delta$, and consequently to 
lead to more stringent constraints on $\gamma$. 

In order to study the implication of this assumption, let us consider
the isospin structure of the amplitudes for the decays $B \to \bar{D} K$
and $B \to D K$. Since the transition operators for 
$\bbar \to \cbar u \sbar$ and $\bbar \to \ubar c \sbar$ are both pure 
$\Delta I = 1/2$, these
processes can be described in terms of two {\it independent} pairs of 
complex
amplitudes, corresponding to the two final mesons being in $I=0$ and 
$1$ states  \cite{Koide,Desh}. Thus, we have for $B \to \bar{D}K$ from 
$\bbar \to \cbar u \sbar$ 
$$
A(B^+ \to \Dbar K^+) = \Abar_1 e^{i\bar{\delta_1}} = \Tbar + \Cbar~~,
$$
$$
A(B^0 \to D^- K^+) = {1\over 2} \Abar_1 e^{i\bar{\delta_1}}  - 
{1\over 2} \Abar_0 e^{i\bar{\delta_0}} = \Tbar~~,
$$
\beq\label{isoDbar}
A(B^0 \to \Dbar K^0) = {1\over 2} \Abar_1 e^{i\bar{\delta_1}}  + 
{1\over 2} \Abar_0 e^{i\bar{\delta_0}} = \Cbar~~,
\eeq
and for $B \to D K$ from $\bbar \to \ubar c \sbar$
$$
A(B^+ \to D^0 K^+) = [{1\over 2} A_1 e^{i\delta_1}  + 
{1\over 2} A_0 e^{i\delta_0}] e^{i\gamma} = C + A~~,
$$
$$
A(B^+ \to D^+ K^0) = [{1\over 2} A_1 e^{i\delta_1}  - 
{1\over 2} A_0 e^{i\delta_0}] e^{i\gamma} = -A~~,
$$
\beq\label{isoD}
A(B^0 \to D^0 K^0) = A_1 e^{i\delta_1} e^{i\gamma} = C~~.
\eeq
We note that there are four independent CP-conserving phases describing 
in general the dominantly inelastic rescattering in the two pairs of 
$I=0$ and 
$1$ channels. In Ref.~\cite{Xing}
(and also in \cite{Desh}) the corresponding phases in $B \to \bar{D} K$ 
and 
in $B \to D K$ were assumed to be equal, $\bar{\delta_i} = 
\delta_i,~i=1,2$. We do not expect this to be the case in general, 
owing to the different hadronic dynamics following the distinct $\bbar 
\to \cbar u \sbar$ and 
$\bbar \to \ubar c \sbar$ quark subprocesses as described in the next paragraph. 

The right-hand-sides of Eqs.~(\ref{isoDbar}) and (\ref{isoD}) consist of 
{\it equivalent} expressions in terms of a graphical description of 
amplitudes, where overall signs follow from a specific phase convention 
for meson states \cite{SU3BR}.
$\Tbar$ is a {\it tree} amplitude involving the subprocess $\bbar \to 
\cbar u \sbar$ in which the $u \sbar$ produced by the weak current 
materializes into a single meson in a color-favored manner. $\Cbar (C)$ 
is a {\it color-suppressed} amplitude for $\bbar \to \cbar u \sbar$ 
($\bbar \to \ubar c \sbar$), where the $u \sbar$ ($c \sbar$) pairs 
produced by the weak current end up in different mesons;  and $A$ 
describes {\it annihilation} of the $\bbar$ and the $u$ in a decaying 
$B^+$ into a weak current, which then materializes into a pair of mesons. 
The processes $B \to \bar{D} K$ are 
written in terms of $\Tbar$ and $\Cbar$, while $B \to D K$ are given by 
$C$ and $A$.

The assumption $A=0$ implies equalities between the magnitudes and 
phases of the two isospin amplitudes in $B \to D K$, $A_1 = A_0$, 
$\delta_1 = \delta_0$, and consequently 
\beq
r = {A_1 \over \Abar_1} = \vert{C \over \Tbar + \Cbar}\vert~~,
~~~~~~\delta=\delta_1 - \bar{\delta}_1~~.
\eeq
Clearly $A_1/\Abar_1$ and $\delta_1 - \bar\delta_1$ are two independent parameters.
They remain independent also when assuming factorization for the ratios
of amplitudes $C/\Tbar$ and $\Cbar/\Tbar$.
To calculate $r$ using generalized factorization \cite{Neubert} for color-allowed ($\Tbar$) {\it and} color-suppressed ($\Cbar, C$)
amplitudes, one would need information about the relative
strong phase between $\Tbar$ and $\Cbar$. In the absence of information
about the interference between the two terms,
one can obtain an approximate estimate by disregarding the smaller 
$\Cbar$ contribution. Thus one finds $r \approx \vert C / \Tbar \vert
\approx \vert V^*_{ub}V_{cs}/V^*_{cb}V_{us}\vert(a_2/a_1)\approx 0.1$, 
where a value of 0.4 is taken for the CKM ratio \cite{Drell} and the 
color suppression factor $a_2/a_1 \approx 0.26$ is taken from a study 
of $B \to \bar{D} \pi$ decays \cite{Browder}. On the other hand, 
$\delta$ remains arbitrary. 

This situation is to be contrasted with the arguments presented in 
\cite{Xing}, where $\delta_i = \bar{\delta}_i$ was assumed, and 
consequently $r$ and $\delta$ were found to be related to each other 
when factorization 
was assumed. We stress again that, in general, no such phase relation 
is expected. The analysis of \cite{Xing} is clearly expected to hold 
in the 
limit in which all strong phases are assumed to vanish; however in 
reality these phases could be sizable. In the case of vanishing phase differences, Eqs.~(\ref{R12}) imply (without assuming $A=0$) a simple 
relation
\beq
\cos\gamma = {R_1 - R_2 \over 4r}~~,
\eeq
which would permit a determination of $\gamma$ from $R_1$ and $R_2$ 
once $r$ is known.  

Final state phases in $B \to \bar{D} K$ can be studied experimentally. 
Similar studies were carried out in $B \to \bar{D} \pi$, and 
an upper limit on the corresponding final state phase-difference was 
obtained at a level of $20^{\circ}$ \cite{Nelson}. The three amplitudes 
in Eqs.~(\ref{isoDbar}) obey a triangle relation \cite{SU3BR}, 
\beq\label{TR1}
A(B^0 \to D^- K^+) +  A(B^0 \to \Dbar K^0)  = A(B^+ \to \Dbar K^+)~~,
\eeq
shown in Fig. 1, where amplitudes are denoted by the flavor of $B$ 
and $D$. 
The dashed-dotted line (of length $\bar{A}_0 /2$) divides 
$A(B^+ \to \Dbar K^+)$ into two equal segments 
each of length $\bar{A}_1 /2$ and forms an angle $\bar{\delta}_0 - \bar{\delta}_1$ with this amplitude. The rate of $B^0 \to 
\Dbar K^0$ would require tagging the flavor of the initial neutral 
$B$ to avoid interference with $\Bbar^0 \to \Dbar \Kbar$. 
%MG some change and addition
(Self-tagged $B \to \bar{D} K^*$ are advantageous in this respect).
A similar study of $\delta_0 - \delta_1$, using the triangle formed by 
the 
three amplitudes of Eqs.~(\ref{isoD}), is inhibited by the difficulty 
of measuring the amplitude of $B^+ \to D^0 K^+$, where $D^0$ is 
identified by a Cabibbo-allowed decay. 
This amplitude interferes 
strongly with $B^+ \to 
\Dbar K^+$, where $\Dbar$ decays to the same state in a doubly Cabibbo-suppressed manner. With a very large number of $B$ mesons 
produced 
in dedicated hadronic $B$ production experiments \cite{Stone}, 
the magnitude of this amplitude can be determined by
observing two different neutral $D$ meson final states \cite{ADS}.

Evidence for a small final state phase difference $\delta_0 - 
\delta_1$ in $B \to D K$ can also be obtained 
by an experimental confirmation of a very small rate for $B^+ \to 
D^+ K^0$ given by $\vert A\vert^2$. Assuming a characteristic 
hierarchy of amplitudes
$\vert A \vert \sim 0.2 \vert C\vert$ \cite{SU3BR, anni}, one 
expects with no rescattering $\vert A(B^+ \to D^+ K^0) \vert  
\sim 0.2 \vert A(B^+ 
\to D^0 K^+) \vert \sim 0.02 \vert A(B^+ \to \Dbar K^+) \vert$. 
Consequently, using the measured rate for $B^+ \to \Dbar K^+$ 
\cite{CLEO}, 
one estimates $\b(B^+ \to D^+ K^0) \sim 10^{-7}$. Such a rate 
measurement (or an upper limit at this level) is attainable in 
an upgrade version of 
CESR \cite{Wein}, PEP-II \cite{PEP}, or KEK-B \cite{KEK}, as long 
as 300 million $B^+B^-$ pairs can be produced, and in proposed 
hadronic experiments \cite{Stone}. A much larger rate would indicate 
significant rescattering effects. These effects could occur through 
much less suppressed intermediate states such as $D^+_s \pi^0$ and 
$D^+_s \eta(\eta')$ \cite{BGR}, the branching ratios of which are 
expected to be larger than the above by a factor of about 
$(0.2)^{-4}\approx 600$ according to the same hierarchy.

Assuming that $A$ is small and can be neglected relative to $C$ 
(i.e. that a branching ratio $\b(B^+ \to D^+ K^0) \sim 10^{-7}$ 
or smaller is measured), one can gain knowledge of $\gamma$ by 
supplementing information from color-allowed $B^{\pm} \to 
D K^{\pm}$ with rates of isospin-related neutral $B$ decays 
\cite{HoKo}. Using the isospin relation Eq.~(\ref{TR1}) 
and the approximate equality (neglecting $A$)
\beq
A(B^+ \to D^0 K^+) \approx A(B^0 \to D^0 K^0)~~,
\eeq
one finds
\beq\label{TR2}
A(B^0 \to D^- K^+) +  \s A(B^0 \to D^0_1 K^0)  \approx \s A(B^+ 
\to D^0_1 K^+)~~.
\eeq
A similar triangle relation 
\beq\label{TR3}
A(\Bbar^0 \to D^+ K^-) +  \s A(\Bbar^0 \to D^0_1 \Kbar)  \approx 
\s A(B^- \to D^0_1 K^-)~~,
\eeq
holds for the charge conjugate amplitudes, obtained from 
Eqs.~(\ref{isoDbar}) and (\ref{isoD}) by replacing $\gamma$ 
with $-\gamma$. (The triangle (11) is 
unchanged by charge conjugation).
The three triangles (\ref{TR1})(\ref{TR2}) and (\ref{TR3}), shown in 
Fig.~1, share a common base
$A(B^0 \to D^- K^+) = A(\Bbar^0 \to D^+ K^-)$ and are fixed, up to 
discrete ambiguities, by seven rate measurements. The angle between 
the two broken 
lines connecting the apex of (\ref{TR1}) to the apexes of the two 
other triangles is $2\gamma$. 
%MG 
Measurement of the lengths of these two lines, which requires very 
high statistics to separate Cabibbo-allowed $D^0$ decays in $B^+ 
\to D^0 K^+$ 
from doubly Cabibbo-suppressed $\Dbar$ decays in $B^+ \to \Dbar K^+$ \cite{ADS}, would provide self-consistency checks.

This method of measuring $\gamma$ from $B \to D K$, or preferably 
$B \to D K^*$ \cite{HoKo},
demonstrates the power of neglecting the annihilation amplitude. It 
involves a relatively large number of processes, none of which is 
suppressed 
by both $V_{ub}$ and color. All the measured rates are governed by 
$\vert V_{cb}V_{us} \vert$. The rates of the three color-suppressed 
neutral 
$B$ decays to neutral $D$ and $K$ mesons are expected to be smaller 
than the other four rates. Using $\vert \Cbar / \Tbar \vert \sim 0.2$ \cite{SU3BR}, 
one estimates $\b(B^0 \to \Dbar K^0) \sim 10^{-5}$. The efficiency for observing $D^0$ CP-eigenstates is somewhat lower than the one for 
detecting $D^0$ or $\Dbar$, at a level of a few percent \cite{Soffer}. 
With 300 million $B^0\bar{B}^0$ pairs, one may expect the precision in measurements of the smaller amplitudes to be at a level of $10\%$. 
(We disregard a tagging efficiency, since the same analysis can be 
applied 
to self-tagged $B \to D K^*$). The errors on the other sides of the 
triangles are smaller. The neglect of the amplitude $A$ relative to $C$ contributes to a larger error, at a level of $20\%$ (assuming a 
hierarchy 
$\vert A/C \vert \sim 0.2$ \cite{SU3BR}), and is the main source 
for the error in $\gamma$.
Carrying out the program of Ref.~\cite{ADS} to measure also the smaller 
dotted lines representing $A(B^+ \to D^0 K^+)$ and $A(B^- \to 
\Dbar K^-)$
could reduce this error and resolve the discrete amibiguities 
in $\gamma$. 
We stress again that neutral $B$ decays to neutral $D$ and $K$ 
mesons must be flavor-tagged. This can be avoided by studying the 
corresponding decays $B \to D K^*$, in which the charged $K$ 
from $K^* \to K \pi$ tags the flavor of $B$. 

In summary, we studied the implications on final state phases of a 
negligible annihilation contribution in $B \to D K$ decays. Contrary 
to a claim in \cite{Xing}, we showed that this assumption 
does not lead to any further constraint on $\gamma$ from 
color-allowed $B^{\pm} \to D K^{\pm}$ beyond the ones obtained in 
\cite{MG}. On the other hand, an application 
of this assumption was demonstrated in which $\gamma$ can be 
determined from charged and neutral $B$ decays to $D K$ or $D K^*$ 
states. A test of a sufficiently small annihilation amplitude was 
proposed in $B^+ \to D^+ K^0$ or $B^+ \to D^+ K^{*0}$ requiring 
branching ratios of about $10^{-7}$ or smaller. Conversely, an 
observation of these decays (and their charge 
conjugates) with considerably larger branching ratios would provide 
an early warning of nonnegligible rescattering effects.
                      
\bigskip
%\bigskip
%\centerline{\bf  ACKNOWLEDGEMENTS}
%\bigskip
One of us (M. G.) is grateful to the Fermilab Theoretical Physics Group 
for 
its kind hospitality. We wish to thank J-H. Jang and P. Ko for a very
valuable communication. This work was supported in part by the United 
States  - Israel Binational Science Foundation under Research Grant 
Agreement 94-00253/2, and by the the United-States Department of Energy 
under Contract No. DE FG02 90ER40560. 
\medskip

% Journal and other miscellaneous abbreviations for references
% Phys. Rev. D style
\def \ajp#1#2#3{Am.~J.~Phys.~{\bf#1}, #2 (#3)}
\def \apny#1#2#3{Ann.~Phys.~(N.Y.) {\bf#1}, #2 (#3)}
\def \app#1#2#3{Acta Phys.~Polonica {\bf#1}, #2 (#3)}
\def \arnps#1#2#3{Ann.~Rev.~Nucl.~Part.~Sci.~#1 (#2) #3}
\def \cmp#1#2#3{Commun.~Math.~Phys.~{\bf#1}, #2 (#3)}
\def \ib{{\it ibid.}~}
\def \ibj#1#2#3{~{\bf#1}, #2 (#3)}
\def \ijmpa#1#2#3{Int.~J.~Mod.~Phys.~A {\bf#1}, #2 (#3)}
\def \ite{{\it et al.}}
\def \jmp#1#2#3{J.~Math.~Phys.~{\bf#1}, #2 (#3)}
\def \jpg#1#2#3{J.~Phys.~G {\bf#1}, #2 (#3)}
\def \mpla#1#2#3{Mod.~Phys.~Lett.~A {\bf#1}, #2 (#3)}\def \ib{{\it 
ibid.}~}
\def \ibj#1#2#3{~{\bf#1}, #2 (#3)}
\def \ijmpa#1#2#3{Int.~J.~Mod.~Phys.~A {\bf#1}, #2 (#3)}
\def \ite{{\it et al.}}
\def \jmp#1#2#3{J.~Math.~Phys.~{\bf#1}, #2 (#3)}
\def \jpg#1#2#3{J.~Phys.~G {\bf#1}, #2 (#3)}
\def \mpla#1#2#3{Mod.~Phys.~Lett.~A {\bf#1}, #2 (#3)}
\def \nc#1#2#3{Nuovo Cim.~{\bf#1}, #2 (#3)}
\def \npb#1#2#3{Nucl.~Phys. B~{\bf#1}, #2 (#3)}
\def \pisma#1#2#3#4{Pis'ma Zh.~Eksp.~Teor.~Fiz.~{\bf#1}, #2 (#3) [JETP
Lett. {\bf#1}, #4 (#3)]}
\def \pl#1#2#3{Phys.~Lett.~{\bf#1}, #2 (#3)}
\def \plb#1#2#3{Phys.~Lett.~B #1 (#2) #3}
\def \pr#1#2#3{Phys.~Rev.~{\bf#1}, #2 (#3)}
\def \pra#1#2#3{Phys.~Rev.~A {\bf#1}, #2 (#3)}
\def \prd#1#2#3{Phys.~Rev.~D #1 (#2) #3}
\def \prl#1#2#3{Phys.~Rev.~Lett.~#1 (#2) #3}
\def \prp#1#2#3{Phys.~Rep.~{\bf#1}, #2 (#3)}
\def \ptp#1#2#3{Prog.~Theor.~Phys.~{\bf#1}, #2 (#3)}
\def \rmp#1#2#3{Rev.~Mod.~Phys.~{\bf#1}, #2 (#3)}
\def \rp#1{~~~~~\ldots\ldots{\rm rp~}{#1}~~~~~}
\def \stone{{\it B Decays}, edited by S. Stone (World Scientific,
Singapore, 1994)}
\def \yaf#1#2#3#4{Yad.~Fiz.~{\bf#1}, #2 (#3) [Sov.~J.~Nucl.~Phys.~{\bf #1},
#4 (#3)]}
\def \zhetf#1#2#3#4#5#6{Zh.~Eksp.~Teor.~Fiz.~{\bf #1}, #2 (#3) [Sov.~Phys.
- JETP {\bf #4}, #5 (#6)]}
\def \zpc#1#2#3{Zeit.~Phys.~C {\bf#1}, #2 (#3)}

\newpage

\begin{figure}
\centerline{\epsfysize = 6 in \epsffile{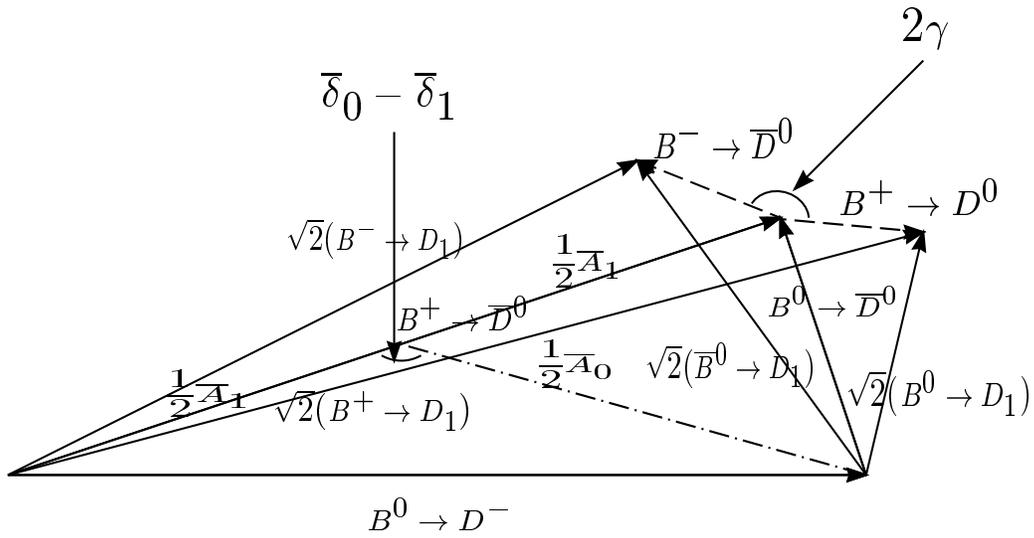}}
\caption{Three triangles representing
Eqs.~(11)(13) and (14) for $B \to D K$ amplitudes. Amplitudes are denoted by the flavor of $B$ and $D$. Dashed-dotted line 
divides $A(B^+ \to \Dbar K^+)$ into two equal segments.}
\end{figure}

\end{document}